\documentstyle[12pt,bezier]{article}
\def\beq{\begin{equation}}
\def\eeq{\end{equation}}

\def\ap#1#2#3 {Ann. Phys. (NY) {\bf#1} (19#2) #3}
\def\apj#1#2#3 {Astrophys. J. {\bf#1} (19#2) #3}
\def\apjl#1#2#3 {Astrophys. J. Lett. {\bf#1} (19#2) #3}
\def\app#1#2#3 {Acta. Phys. Pol. {\bf#1} (19#2) #3}
\def\ar#1#2#3 {Ann. Rev. Nucl. Part. Sci. {\bf#1} (19#2) #3}
\def\cpc#1#2#3 {Computer Phys. Comm. {\bf#1} (19#2) #3}
\def\err#1#2#3 {{\it Erratum} {\bf#1} (19#2) #3}
\def\ib#1#2#3 {{\it ibid.} {\bf#1} (19#2) #3}
\def\jmp#1#2#3 {J. Math. Phys. {\bf#1} (19#2) #3}
\def\ijmp#1#2#3 {Int. J. Mod. Phys. {\bf#1} (19#2) #3}
\def\jetp#1#2#3 {JETP Lett. {\bf#1} (19#2) #3}
\def\jpg#1#2#3 {J. Phys. G. {\bf#1} (19#2) #3}
\def\mpl#1#2#3 {Mod. Phys. Lett. {\bf#1} (19#2) #3}
\def\nat#1#2#3 {Nature (London) {\bf#1} (19#2) #3}
\def\nc#1#2#3 {Nuovo Cim. {\bf#1} (19#2) #3}
\def\nim#1#2#3 {Nucl. Instr. Meth. {\bf#1} (19#2) #3}
\def\np#1#2#3 {Nucl. Phys. {\bf#1} (19#2) #3}
\def\pcps#1#2#3 {Proc. Cam. Phil. Soc. {\bf#1} (#2) #3}
\def\pl#1#2#3 {Phys. Lett. {\bf#1} (19#2) #3}
\def\prep#1#2#3 {Phys. Rep. {\bf#1} (19#2) #3}
\def\prev#1#2#3 {Phys. Rev. {\bf#1} (19#2) #3}
\def\prl#1#2#3 {Phys. Rev. Lett. {\bf#1} (19#2) #3}
\def\prs#1#2#3 {Proc. Roy. Soc. {\bf#1} (19#2) #3}
\def\ptp#1#2#3 {Prog. Th. Phys. {\bf#1} (19#2) #3}
\def\ps#1#2#3 {Physica Scripta {\bf#1} (19#2) #3}
\def\rmp#1#2#3 {Rev. Mod. Phys. {\bf#1} (19#2) #3}
\def\rpp#1#2#3 {Rep. Prog. Phys. {\bf#1} (19#2) #3}
\def\sjnp#1#2#3 {Sov. J. Nucl. Phys. {\bf#1} (19#2) #3}
\def\spj#1#2#3 {Sov. Phys. JEPT {\bf#1} (19#2) #3}
\def\spu#1#2#3 {Sov. Phys.-Usp. {\bf#1} (19#2) #3}
\def\zp#1#2#3 {Zeit. Phys. {\bf#1} (19#2) #3}
\def\as{\alpha_s}
\def\ams{\alpha_s^{\overline {MS}}}
\begin{document}
\begin{titlepage}
\begin{center}
{\Large \bf Theoretical Physics Institute \\
University of Minnesota \\}  \end{center}
\vspace{0.3in}
\begin{flushright}
TPI-MINN-95/1-T \\
UMN-TH-1326-95 \\
February 1995
\end{flushright}
\vspace{0.4in}
\begin{center}
{\Large \bf Precision determination of $\alpha_s$ and $m_b$ from QCD sum
rules for $b \overline b$. \\}
\vspace{0.2in}
{\bf M.B. Voloshin  \\ }
Theoretical Physics
Institute, University of Minnesota \\ Minneapolis, MN 55455 \\ and \\
Institute of Theoretical and Experimental Physics  \\
                         Moscow, 117259 \\
\vspace{0.3in}
{\bf   Abstract  \\ }
\end{center}
The QCD sum rules for moments of production cross section
of $b \overline b$ states in
$e^+\,e^-$ annihilation are extremely sensitive to the values of $m_b$ and
$\alpha_s$ for moments of large order $n$. This enables one to extract from
the existing data  on $\Upsilon$ resonances the values of these parameters
with a high precision by using a non-relativistic expansion in $1/n$.  It is
found that the sum rules fit the data with $\alpha_s^{\overline {MS}} (1 \,
GeV) = 0.336 \pm 0.011$ and $m_b=4827 \pm 7 MeV$, where the estimate of the
errors includes the theoretical uncertainty due to subleading in $1/n$ terms
and the experimental uncertainty of the $e^+\,e^-$
annihilation cross section above the $B \overline B$ threshold.
The found value of
$\alpha_s$, when evolved in two loops up to the $Z$ mass, gives
$\alpha_s^{\overline {MS}} (M_Z) = 0.109 \pm 0.001$. The $b$ quark mass
$m_b$ corresponds to the `on shell' value appropriate for one-loop
perturbative calculations.

\end{titlepage}

\section{Introduction}

Precision tests of the Standard Model, which are becoming possible due
to experimental technique, call for a better precision in understanding
the parameters of the Model as we see it now. In particular the value of
the $b$ quark mass enters as a parameter in the predictions for the
decays of $B$ hadrons, and its understanding is necessary e.g. for a
precise determination of the weak mixing element $V_{cb}$ from
the data on inclusive decays of $B$ hadrons. The value of $\alpha_s$
determined from a global fit to the data of the LEP
experiments at the $Z$ peak: $\alpha_s^{\overline {MS}} (M_Z)=0.125
\pm 0.005 \pm 0.002$ (see e.g. in \cite{vys} and also in
\cite{langacker}) is argued$^{\cite{vys1,shifman}}$ to be
meaningfully higher than what one finds from extrapolation to the $Z$
mass scale of the values of the QCD coupling found  in a number of
analyses at low energies. If
confirmed with further improvement of experimental accuracy and of the
theoretical understanding, this mismatch in the values of $\as$ may
signal a contribution of a new physics to decays of
$Z\,^{\cite{vys1,shifman}}$. It should be noted however, that at present
there exists a variety of low-energy estimates of $\as$, with various
degree of compatibility with the LEP value,
which reflects the present uncertainty in the
extraction of $\alpha_s$ from the low-energy data.
(The most recent and extensive review of various fits of the value of
$\as$ is given by Hinchliffe$^{\cite{hinchliffe1,hinchliffe2}}$.)

The purpose of this paper is to present one more determination of
$\as$ from the low-energy phenomenology, namely from an analysis of the
QCD sum rules for the cross section of production of the $b \overline b$
hadronic states $X_{b \overline b}$ in $e^+\,e^-$ annihilation.
Simultaneously the same analysis yields a precision determination
of the mass parameter $m_b$. For the states $X_{b \overline b}$ the QCD sum
rules${\cite{sr}}$ relate the integral moments of the physically measured
quantity  $R_b=\sigma(e^+\,e^- \to X_{b \overline b})/
\sigma(e^+\,e^- \to \mu^+\,\mu^-)$ of the form
$\int ds \,R_b(s)/s^{n+1}$ to
theoretically calculable derivatives of the vacuum polarization by the
vector current $({\overline b} \, \gamma_\mu \, b)$. For high enough $n$
the moments are saturated by the lowest vector resonances: $\Upsilon$'s,
and are essentially not sensitive to the uncertainty of the cross
section in the continuum above the $B \overline B$ threshold.

It has been noticed long ago$^{\cite{sr,v1}}$ that for large $n$ the
theoretical calculation of the moments within the perturbation theory
contains as a parameter $\as\,\sqrt{n}$ rather than $\as$, which in a
dispersive calculation corresponds to dominance of the near-threshold
quark-antiquark dynamics at typical velocity $v \approx 1/\sqrt{n}$.
So that $\as \, \sqrt{n} \approx \as/v$ is the familiar Coulomb
parameter. Therefore at large $n$ the Coulomb effects should be
explicitly summed up. On the other hand, this behavior implies that high
moments are very sensitive to the value of $\as$ and thus this value can
be extracted with high accuracy, even though the experimental input is
not very precise. Also, by dimension, the $n$-th moment depends on
$m_b^{2n}$, which explains the high sensitivity at large $n$ to the
quark mass.

The limiting factor in considering moments with high $n$ is the growth
of the relative contribution of the non-perturbative terms. The first
one, appearing in the sum rules for heavy quarks, is
proportional to the so-called gluon
vacuum condensate $\langle \as \, G_{\mu \nu}^2 \rangle \, ^{\cite{svz}}$.
This term grows approximately as $n^3$ relative to the perturbative
one$^{\cite{svz,v1}}$. However in the case of the sum rules for the $b
\overline b$ states its magnitude is still within about 1\% at
$n=20$ (and becomes rapidly more important at higher $n$).
In what follows the range of moments from $n=8$ to $n=20$ will be
considered, where the non-perturbative contribution can be safely
neglected and systematically the leading in $1/n$ approximation will be
used, which, in particular, allows summation of the Coulomb terms
$(\as\,\sqrt{n})^k$.

An analysis of the sum rules for the $b \overline b$ states along these
lines was done quite some time ago$^{\cite{v1}}$ using then available
data and a higher range of $n$: from $n \approx 25$ to $n \approx 40$,
with taking into
account the first non-perturbative term. Those estimates resulted in
evaluating the mass parameter $m_b=4.80 \pm 0.03 GeV$ and in an estimate
of $\as$: $\as=0.30 \pm 0.03$ at a momentum scale of order 1 GeV. Here
the analysis is refined by calculating
the effects of running of the coupling $\as$ thereby precisely
specifying the scale for $\as$ as a function of $n$.
Also an uncertainty due to the next
term in the $1/n$ expansion is estimated, which allows to consider a
lower range of $n$, where the results are not affected by an uncertainty
in the value of the gluon condensate.

One can present several reasons for this method of determining $\as$
being one of the most, if not the most, reliable. Firstly,
it is fully justified within the short-distance QCD and does not
rely on assumptions about local duality, as one has to, when considering
rates at fixed energy, like, say, in the annihilation of heavy quarkonia
into gluons, or even in more traditional processes like the total cross
section of $e^+\,e^-$ annihilation into hadrons at fixed energy, or the
total hadronic decay rate of the $\tau$ lepton\footnote{The later method
of determining $\as$ (see e.g. in \cite{pich}) was recently criticized
by Shifman$^{\cite{shifman}}$ on the basis of possible corrections to
local duality in the timelike domain.}. Secondly, since the parameter
$\as \sqrt{n}$ in the considered range of $n$ is of order one, the
effects of the coupling are not just small corrections but in fact
are dominating. Therefore the value of the coupling can be determined
with a high precision. Thirdly, the moments with sufficiently high $n$
are only very weakly sensitive to the poorly known cross section above
the $B \overline B$ threshold. Thus even quite conservative assumptions
about this cross section are sufficient for performing the analysis with
an acceptable accuracy.

\section{Sum rules}

The QCD sum rules$^{\cite{sr}}$ under discussion arise from
considering the vacuum
polarization operator $P_b(q^2)$ induced by the electromagnetic current
$j_\mu = Q_b\, (\overline b \, \gamma_\mu \, b)$ of the $b$ quarks:
\beq
P_b(q^2)={{-i} \over {3\,q^2}} \int \, e^{iqx} \, \langle 0 | T\,\bigl(
j_\mu(x)\,j_\mu(0) \bigr ) \,| 0 \rangle
\label{pb}
\eeq
with $Q_b=-1/3$ being the electric charge of $b$ quark. The imaginary
part of $P_b(s)$ is related to the physically measurable quantity
$R_b(s)$: Im$P_b(s)=R_b(s)/12\pi$. Therefore using dispersion relation
one can write the $n$-th derivative of $P_b(q^2)$ at $q^2=0$ in terms of
the $n$-th integral moment of $R_b(s)$:
\beq
12\pi^2 \, \left. \left ( {d \over {dq^2}} \right )^n \, P_b(q^2) \right
|_{q^2=0} = \int \, R_b(s)\, {{ds} \over {s^{n+1}}}~~~,
\label{dn}
\eeq
where the integral runs over all physical values of $s$ where the cross
section for production of states containing the $b \overline b$ quark
pair in $e^+\,e^-$ annihilation is non-zero, i.e. it includes the
$\Upsilon$ resonances and the continuum above the $B \overline B$
threshold. On the other hand the same derivatives as in eq.(\ref{dn})
can be calculated by methods of the short-distance QCD. In particular in
perturbation theory this can be done by the usual calculation of Feynman
graphs for $P_b(q^2)$, which  in any finite order in $\as$
is equivalent to writing those derivatives
in terms of the formal integrals with the value of $R_b(s)$ given by the
perturbation theory in that order in $\as$: $R_b^{pt}(s)$. In this way
one arrives at the equations
\beq
\int \, R_b(s)\, {{ds} \over {s^{n+1}}} =
\int \, R_b^{pt}(s)\, {{ds} \over {s^{n+1}}} + non\!-\!pert.~terms
\label{sr}
\eeq
which express the QCD sum rules in this particular case. The left hand
side of the sum rules contains a physically measurable quantity, while
the right hand side is calculable theoretically including the first
non-perturbative terms$^{\cite{sr,svz}}$.

\subsection{Non-relativistic limit in the sum rules}

One can notice that at large $n$ the weight function $1/s^{n+1}$ rapidly
decreases with energy, so that the integral is dominated by the lowest
states of $b \overline b$. For the perturbation theory integral in
eq.(\ref{sr}) this implies that at large $n$ the integral is dominated
by the states which are close to the quark-antiquark threshold $s_0
=4m_b^2$. Defining $E$ as the energy counted from the quark threshold,
i.e. $s=(2m_b+E)^2$, one can write
\beq
\int \, R_b(s)\, {{ds} \over {s^{n+1}}}=
\int \, R_b\, {{2\,(2m_b+E)\,dE} \over {(2m_b+E)^{2n+2}}}=
{1 \over {(4 m_b^2)^n}}
\int \, R_b\, \exp \left ( -{E \over m_b} n \right )\,
{dE  \over 2m_b}\, \left ( 1+ O \left( {1 \over n} \right) \right)~,
\label{nr}
\eeq
thus effectively replacing the power weight function by an exponential.
Moreover, it is clear that the integral is determined by the range of
energy where $|E|/m_b \sim 1/n$, i.e. by the non-relativistic region
near the threshold. If one also recalls the relation of the energy to
the velocity $v$ of each of the quarks (in the c.m. system):
$E=m_b\,v^2$, one concludes that the relevant range of velocity is given
by $v \sim n^{-1/2}$, so that the $1/n$ expansion of the integrals
in the sum rules is equivalent to a non-relativistic expansion in
$v^2\,^{\cite{v2}}$.

Furthermore, in the non-relativistic region the spectral density of the
electromagnetic current can be expressed through the spectral
density of the non-relativistic operator $\delta({\rm \bf r})$,
which is convenient to represent in terms of the matrix element $\langle
0 | (H-E)^{-1}|0 \rangle$  of the Green's function $\langle
{\rm \bf x} | (H-E)^{-1} | {\rm \bf y} \rangle$ with the
non-relativistic Hamiltonian $H\,^{\cite{v2}}$:

\beq
R_b ((2m_b+E)^2)= \left ( 1- {{16 \, \as} \over {3 \pi}} \right )\,
{{18 \pi \, Q_b^2} \over {m_b^2}}\, {\rm Im} \langle
0 | (H-E)^{-1}|0 \rangle~.
\label{fg}
\eeq
Here is also included the radiative correction, which comes from
distances of order $1/m_b$ and represents a finite renormalization of the
electromagnetic current at the threshold. This correction will be
further discussed and quantified at the end of this section.

Substituting the expression (\ref{fg}) into the exponential integral in
eq.(\ref{nr}) one finds a relation$^{\cite{v1}}$
of the integral moments to the matrix
element $K(\tau)=\langle 0 | \exp (-H\, \tau) |0 \rangle$ of the
Euclidean time propagation operator $\exp (-H \, \tau)$ at $\tau=n/m_b$:

\beq
\int \, R_b(s)\, {{ds} \over {s^{n+1}}} =
\left ( 1- {{16 \, \as} \over {3 \pi}} \right )\,{{18 \pi^2 \, Q_b^2}
\over {(4 m_b^2)^n \, m_b^3}} \, K\left( {n \over m_b} \right) \,
\left ( 1+ O \left( {1 \over n} \right) \right)~.
\label{rk}
\eeq
The latter representation in terms of $K(\tau)$ enables a detailed
analysis of the effects of quark interaction in the non-relativistic
domain both at the perturbative and the non-perturbative level.

\subsection{Summing Coulomb effects}

In the lowest order of perturbation theory, i.e. for free quarks the
propagation function has the familiar form (see e.g. in \cite{fh},
and it should also be taken into account that the reduced mass in
the $b \overline b$ system is $m_b/2$):
\beq
\langle {\rm \bf x} | \exp (-H_0 \, \tau) | {\rm \bf y} \rangle = \left (
{m_b \over {4\pi \, \tau}} \right )^{3/2} \, \exp \left ( -{m_b \over
{4\,\tau}} ({\rm \bf x} - {\rm \bf y})^2 \right )~,
\label{k0xy}
\eeq
so that
\beq
K_0(\tau)=\left ( {m_b \over {4\pi \, \tau}} \right )^{3/2} ~,
\label{k0}
\eeq
which reproduces the large $n$ limit of the moments of $R_b(s)$
calculated from the simple loop of Fig.1.

The QCD interaction of quarks via gluons in the leading in $v^2 \sim
1/n$ approximation reduces to a Coulomb-like potential
\beq
V(r)=-{{4 \as} \over {3 r}}~.
\label{pot}
\eeq
The perturbation theory expansion in this potential generates an
expansion for $K(\tau)$ in powers of the parameter $\as \,\sqrt{m_b
\tau}=\as \, \sqrt{n}$. Since we are aiming at the region of $n$
where this parameter is of order one, the interaction (\ref{pot})
should be taken into account exactly rather than perturbatively.
In practice this amounts to finding the propagation function
$K(\tau)$ in the Coulomb problem. In terms of Feynman graphs this is
equivalent to summing the diagrams of Fig.2 in the non-relativistic
region. The time-dependent propagator in the Coulomb field can be
found$^{\cite{v1,v3}}$
by the inverse Laplace transform of the energy-dependent Green's
function at negative energy $E=-k^2/m_b$, which can be derived from a
solution of the Schr\"odinger equation in terms of the singular
at $z=0$ confluent hypergeometric function $U(a, \, b; \, z)$:
\beq
\langle r | \left(H+{k^2 \over m_b}\right)^{-1} | 0 \rangle =
{{m\,k} \over {2 \pi}}
\, e^{-kr} \,\Gamma(1-\lambda)\,U (1-\lambda,\,2, \, 2kr)=
{{m\,k} \over {2 \pi}}\, e^{-
kr} \, \int_0^\infty \, e^{-2krt} \left ( {{1+t} \over t} \right
)^\lambda \, dt~,
\label{sse}
\eeq
where $\lambda=2 m_b \, \as/(3k)$ is the Coulomb parameter for the
potential (\ref{pot}). Setting $r=0$ in the integral representation in
eq.(\ref{sse}) gives a divergent integral. However the divergent term
does not depend on $k$ and thus does not affect the inverse Laplace
transform in the variable $k^2/m$. The result for $K(\tau)$ has the
form$^{\cite{v1}}$ $K(n/m_b)=K_0(n/m_b) \, F(\gamma)$,
where $\gamma=2 \as \,
\sqrt{n}/3$ is the Coulomb parameter in the time-dependent
representation, and
\beq
F(\gamma)=1+2\sqrt{\pi}\gamma + {{2 \pi^2} \over 3} \gamma^2+
4\sqrt{\pi}\, \sum_{p=1}^\infty \left ( {\gamma \over p}\right )^3\,\exp
\left [ \left( {\gamma \over p} \right )^2 \right ]\,
\left[ 1 + {\rm erf} \left ( {\gamma \over p} \right ) \right ]~,
\label{fc}
\eeq
where, as usual, erf$(x)=(2/\sqrt{\pi}) \, \int_0^x \, \exp(-t^2)\, dt$.
One can readily see that the series in eq.(\ref{fc}) is well converging,
so that this expression can be used for practical calculation. The
function $F(\gamma)$ sums all the terms of the form $(\as \,
\sqrt{n})^k$ in the time dependent propagator and thus in the sum rules.
One can also easily recognize that in the formal limit $n \to \infty$
the function $F(\gamma)$ is determined by the first term in the sum in
eq.(\ref{fc}): $F(\gamma) \to 8 \sqrt{\pi} \gamma^3 \, \exp(\gamma^2)$
which is exactly the contribution of the lowest Coulomb bound
state\footnote{It should be noticed however that at finite $\gamma$ each
$p$-th term in the sum in eq.(\ref{fc}) receives contribution both from the
$p$-th Coulomb bound state and from the continuum above the threshold.}.
In the analysis of the sum rules for the $b \overline b$ system we will
be restricted by the region, where $\gamma$ is of order 1. In connection
with this it can be noticed that numerically the function $F$ is large
in this region: $F(1)=49.1243\ldots$, which explains the high sensitivity
of the sum rules to the value of $\as$.

\subsection{Effects of running $\as$}

In the previous discussion the effects of running of the renormalized
coupling $\as$ were ignored. These have to be included in order to be
able to relate the moments of $R_b$ to the value of $\as$ at a specified
scale. Naively, one would estimate that at a velocity $v$ the typical
momentum flowing through the Coulomb gluons in the graphs of Fig.2 is of
order $m_bv$, which in the $\tau$ domain translates into the expectation
that the normalization for $\as$ scales as $\sqrt{m_b/\tau} =
m_b/\sqrt{n}$. However the Coulomb effects substantially modify this
estimate in the region where $\as/v$, or equivalently $\gamma$,
is not small. This can be seen from the simple fact that in the formal
limit $\gamma \to \infty$ the function $K(\tau)$ is determined by the
lowest Coulomb level, which has the intrinsic scale given by the Bohr
momentum $k_B = 2 \as m_b/3$. Therefore one should expect that the
proper normalization scale for $\as$ in the propagation function
$K(\tau)$ is given by $ \sqrt{m_b/\tau} \, h(\gamma)$, i.e.
$K(\tau)=K_0(\tau) \, F(\gamma( \sqrt{m_b/\tau} \, h(\gamma)))$
where the function $h(\gamma)$ describes the Coulomb effects on the
scale and the overall normalization of $h(\gamma)$
depends on the renormalization scheme for
$\as$. In particular the dominance of the lowest Coulomb level at large
$\gamma$ implies that in this limit $h(\gamma) \propto \gamma$.

To quantify the effects of running $\as$ we adopt the $\overline {MS}$
scheme, in which the $\as$ in the interaction potential in eq.(\ref{pot})
is$^{\cite{blm}}$ $\ams(\kappa/r)$ with $\kappa=\exp (-C-5/6)$ and where
$C=0.5772\ldots$ is the Euler constant. When expressed in terms of
$\ams(\mu)$ normalized at a fixed scale $\mu$, the effect of running
$\as$ in the potential (\ref{pot}) is reduced to a modification of the
potential of the form:
\beq
V(r)=-{{4 \ams(\mu)} \over {3r}} - {{4 \as}\over {3 r}} { {b\,\as}
\over {2 \pi}} \ln {{\mu r} \over {\kappa}} + \ldots ~,
\label{mpot}
\eeq
where $b=9$ is the first coefficient in the QCD $\beta$ function. (We
deal here with the typical scales about 1 GeV, so that the appropriate
number of active quark flavors in the $\beta$ function is $n_f=3$.)
On the other hand if the function $K(\tau)$ is written in terms of
$\gamma(\mu)= 2 \ams(\mu) \sqrt{n} /3$, the same first order effect in
the running results in the following modification for the function
$F(\gamma)$:
\beq
F=F(\gamma(\mu))+ \gamma \, {{dF} \over  {d \gamma}} \, { {b \,\as} \over
{2 \pi}} \ln \left ( {{\mu \, \sqrt{n}} \over {h(\gamma) \, m_b}} \right)~.
\label{mfc}
\eeq
Finally, the modification of the Coulomb function $F(\gamma)$ is clearly
the result of the modification of the potential in eq.(\ref{mpot}). To
relate the two expressions it is technically convenient to first find
the modification of the energy-dependent Green's function by the
perturbation of the potential as in eq.(\ref{mpot}) and then to find the
corresponding change of the time-dependent propagator by the inverse
Laplace transform. For the energy-dependent Green's function the
correction has the form:
\beq
\delta \langle 0 |\left ( H+ {k^2 \over m_b} \right )^{-1} | 0 \rangle=
{{4 \as}\over {3 }}\, { {b\,\as} \over {2 \pi}}\,{{m^2 \, k^2} \over \pi}
\int\,dt\, du \, dr \, r \,
e^{-2kr(t+u+1)}\, \left ( {{(1+t)(1+u)} \over {t\,u}}
\right )^\lambda \, \ln {{\mu r} \over {\kappa}}~,
\label{intr}
\eeq
where each of the integrals runs from zero to infinity. The integration
over $r$ can be easily done explicitly, which however is not as simple
for the integration over $t$ and $u$. However the integral can be
expanded in powers of $\lambda$ and transformed into the $\tau$
representation term by term, which generates an expansion of the
modification $\delta F(\gamma)$ of the function $F$ in powers of
$\gamma$. (In doing this transformation one only needs the
transformation rule for a generic term $1/k^p$:
$1/k^p \rightarrow \tau^{p/2-1} /\bigl ( m_b^{p/2}\, \Gamma(p/2)
\bigr)$.) This procedure works for all terms of the expansion of the
integral in eq.(\ref{intr}) in powers of $\lambda$, except for the first
one with $\lambda^0$, which contains a manifestly divergent integral.
However this term can be easily calculated directly in terms of
$K(\tau)$ using the free evolution function in eq.(\ref{k0xy}). This
amounts to calculating the integral
\beq
{{m_b^3} \over {16 \, \pi^2}} \int_0^{\tau} \, {{d\tau_1} \over {
\tau_1^{3/2} \, (\tau-\tau_1)^{3/2}}} \, \int \, dr \, \exp \left ( -\,
{{m_b \, r^2 \, \tau} \over {4 \, \tau_1 \, (\tau-\tau_1)}} \right ) \,
\ln r = {{m_b^2} \over {8 \pi \, \tau}} \, \left[\ln \left ( {1 \over 2}
\sqrt{ \tau \over m_b} \right ) - {C \over 2} \right]~.
\label{k1}
\eeq

Collecting all terms and equating the modification of the function
$F(\gamma)$ in eq.(\ref{mfc}) to that calculated from the perturbation
of the potential in eq.(\ref{mpot}) we finally find the function
$h(\gamma)$ in the following form of expansion in powers of $\gamma$:
\beq
\ln h(\gamma)= \ln (2 \kappa) + {{2 \, \sqrt{\pi}} \over {F^\prime
(\gamma)}} \left ( {C \over 2} + 2\,\sum_{p=1}^\infty \, a_p \, \gamma^p
\right)~,
\label{hge}
\eeq
where $F^\prime=dF/d\gamma$ and the coefficients $a_p$ are given by the
integrals
\beq
a_p=\int_0^1 \int_0^1 \, dy \,dz {{\left\{ \psi(p/2)/2-\psi(2)- \ln
\left [ (1-y) (1-z)/(1-y \, z) \right] \right\} \left ( - \ln yz \right
)^p } \over {p! \, \Gamma(p/2) \, (1-y \, z)^2}}~,
\label{ap}
\eeq
where $\psi(x)=d \ln \Gamma(x)/dx$ and the integration variables $y$ and
$z$ are related to those in eq.(\ref{intr}) by the obvious substitution:
$y=t/(1+t),~z=u/(1+u)$. The integrals (\ref{ap}) for the coefficients
$a_p$ can be successfully calculated numerically. The convergence of the
expansion in eq.(\ref{hge}) is however somewhat slow. In particular, for
calculating $h(g)$ with four digit accuracy at $\gamma=1$ one has to
keep 21 first terms in the expansion. (In the numerical analysis,
described in the next section, were used 30 terms of the expansion.)

Thus we find that the normalization point for $\as$ in the parameter
$\gamma$ in eq.(\ref{fc}) is itself determined by the value of $\gamma$.
Therefore in order to take into account the
running of the coupling constant in one
loop one has to determine the value of $\gamma$ in eq.(\ref{fc}) by
solving the equation
\beq
\gamma=\gamma\left( {m_b \over \sqrt{n}} \, h(\gamma) \right )~,
\label{geq}
\eeq
which can be easily done numerically by iterations.

\subsection{Short-distance radiative correction}
Thus discussed radiative effects arise from distances, which are
parametrically larger than $m_b^{-1}$ by a factor $\sqrt{n}\,h(\gamma)$,
which are the characteristic distances in the non-relativistic Coulomb
dynamics of quarks at a velocity $v \sim 1/\sqrt{n}$. There is however
a radiative effect, which comes from the distances of order $m_b^{-1}$
and is associated with correction to the electromagnetic vertex.
(A discussion of this point in QED can be found in the textbook
\cite{schw}.) To quantify this effect let us consider the graphs shown
in Fig.3 for the calculation in up to the one-loop order of
the spatial part of
the vector current $(\overline b \, \gamma_i \, b)$ exactly at the $b
\overline b$ threshold, i.e. for the quark and the antiquark being at
rest in the c.m. system. (It is only the spatial part of the current
which is non vanishing in the c.m. kinematics.) The sum of the graphs
can be written as
\beq
(\overline b \, \gamma_i \, b) \, \left ( 1+ {\as \over {3\, \pi}} J
\right )~,
\label{fj}
\eeq
where the one-loop correction term $J$ after integrating over the
angular variables can be written in the form of an integral over the
Euclidean $k^2$ of the virtual gluon:
\beq
J=\int_0^\infty \left ( {4 \over {t^{3/2}}}
+ w(t) \right ) \, dt ~,
\label{jint}
\eeq
where $t=k^2/m_b^2$ and the integrand contains the infrared-singular
term $4/t^{3/2}$ and the regular part $w(t)$, for which the explicit
expression is
\beq
w(t)={4 \over {3 \, t^{3/2} \, \sqrt{t+4}}} \, \left ( 6-2\,t +2\,t^2
+t^3 - 3\, \sqrt{t+4} - t^{5/2} \, \sqrt{t+4} \right )~.
\label{wt}
\eeq
The infrared divergence of the singular term corresponds to the
singularity of the first Coulomb correction $\propto \as/v$ at velocity
$v=0$, and this term is in fact the one accounted for in the Coulomb
problem calculation$^{\cite{schw}}$. Therefore the additional radiative
correction is described only by the regular part with $w(t)$.
The integral with the $w(t)$ from eq.(\ref{wt}) can be
readily done:
\beq
\int_0^\infty \, w(t) \, dt =-8~,
\label{intw}
\eeq
which results in the well known expression for the radiative correction
to the matrix element squared, used in eq.(\ref{rk}).
What is important here is that the integral (\ref{intw}) is determined
by virtual momenta of order $m_b$ therefore in contrast with the Coulomb
terms this radiative correction depends on $\as$ normalized at a scale
of order $m_b$. A precise specification of this scale in the $\overline
{MS}$ scheme requires a two loop calculation of the form factor at the
threshold, which to the best of my knowledge is so far absent. In view
of this we have to resort to evaluating the scale within the Brodsky -
Lepage - Mackenzie  (BLM) scheme$^{\cite{blm}}$, which technically
amounts to averaging $\ln t$ with the weight function $w(t)$ determining
the one-loop correction. For this average we find
\beq
\langle {1 \over 2} \, \ln t \rangle \equiv {1 \over 2} \,{{\int_0^\infty
\, w(t)\, \ln t \, dt} \over {\int_0^\infty \, w(t)\, dt}} =
{3 \over 8}~.
\label{lav}
\eeq
By the BLM prescription the normalization point for $\as$ is then given
by $e^{3/8} m_b$ in the $V$ scheme (fixed by the Coulomb potential in
the momentum space) and by $e^{-5/6}\, e^{3/8} m_b = e^{-11/24} m_b
\approx 0.632 \, m_b$ in the $\overline {MS}$ scheme. Therefore we
conclude that within the BLM method the appropriate value of $\as$ in
the short-distance radiative correction in eq.(\ref{rk}) is $\ams(0.632
\, m_b)$.

\section{Numerical analysis}

Summarizing the discussion of the previous section, the large $n$
perturbative
formula for the moments of $R_b(s)$, including the effects of running
coupling and of the $O(\as)$ radiative correction, can be written as
\beq
\int \, R_b(s)\, {{ds} \over {s^{n+1}}} =
\left ( 1- {{16 \, \ams(0.632 m_b)} \over {3 \pi}} \right )\,
{{\sqrt{\pi} \,9\, Q_b^2}
\over {(4 m_b^2)^n \, 4\, n^{3/2}}} \, F\left( \gamma(m_b \,
h(\gamma)/\sqrt{n}) \right) \,
\left ( 1+ O \left( {1 \over n} \right) \right)~,
\label{rf}
\eeq
where the function $F(\gamma)$ is given by eq.(\ref{fc}) and its
argument is found by solving the equation (\ref{geq}) with the function
$h$ determined by eqs.(\ref{hge}) and (\ref{ap}). Further perturbative
corrections to the expression (\ref{rf}) are at least as small as
$O(\as^2)$ with no additional enhancement in $n$. On the other hand the
region of the parameters to be considered is where $\as \, \sqrt{n} =
O(1)$. Thus the uncertainty due to further radiative corrections is
parametrically the same as that from the $1/n$ terms.

Another class of corrections to eq.(\ref{rf}) goes beyond perturbation
theory and is associated with non-perturbative properties of the QCD
vacuum. These corrections grow with $n$, and the first one, due to the
gluon condensate, is well known. In the leading $1/n$ approximation this
correction amounts$^{\cite{v1}}$ to replacing $F(\gamma)$ in
eq.(\ref{rf}) by the expression
\beq
F(\gamma) \left ( 1- \xi(\gamma)\, {n^3 \over m_b^4} \, \langle {{\pi \,
\as} \over {72}} \, G_{\mu \nu}^2 \rangle \right )~,
\label{npt}
\eeq
where the function $\xi(\gamma)$ has a rather
complicated form$^{\cite{v1}}$, which
however can be approximated in the region $\gamma \le 1.5$ by simple
exponent: $\xi(\gamma) \approx \exp(-0.8 \gamma)$. Using this
approximation and the value$^{\cite{svz}}$ of the gluon condensate
$\langle { \as \over \pi} G^2 \rangle \approx 0.012 \, GeV^4$, one can
estimate that the relative magnitude of the non-perturbative term does
not exceed about 1\% for $n \le 20$. For this reason the present
analysis is restricted to moments with $n \le 20$ and the non-
perturbative term is completely ignored.

On the lower end the range of $n$ is restricted by the applicability of
the $1/n$ expansion. Another practical restriction is that the value of
$R_b$ is very poorly known experimentally above the $B \overline B$
threshold. To keep this uncertainty well suppressed in the moments it is
desirable to choose larger values of $n$.

In the present analysis as the experimental inputs are used the masses
and the $e^+\,e^-$ widths with their experimental errors for the four
lowest $\Upsilon$ resonances, as given by the Particle Data
Tables$^{\cite{pdg}}$. In connection with this it can be reminded that
the value of $R(s)$ for a narrow resonance can be written as
\beq
R_{Res}(s)={{9\pi} \over {\overline \alpha}^2} \, \Gamma(Res \to e^+ \,
e^-) \, M_{Res} \, \delta(s-M_{Res}^2)~,
\label{res}
\eeq
where $\overline \alpha$ contains the renormalization of the
electromagnetic $\alpha$ at energy near the mass of the resonance:
\beq
{\overline \alpha}^2 = {\alpha^2 \over {|1-P_{em}(M_{Res}^2)|^2}}
\label{abar}
\eeq
with $P_{em}(q^2)$ being the full electromagnetic vacuum polarization
operator. In the numerical analysis here we use the estimate ${\overline
\alpha}^2 = 1.07 \alpha^2$ at energy in the region of $\Upsilon$
resonances. This correction turns out to be quite essential in
the numerical fits.

As to the cross section above the continuum threshold, i.e. above the
$\Upsilon(4S)$, a conservative assumption is adopted here that the value
of $R_b(s)$ is equal to $(1\pm 0.5)/3$ starting from the mass of
$\Upsilon(4S)$. I believe that the assigned error of 50\% well covers
the possible uncertainties in the contribution of the continuum cross
section to the considered moments.

For a fit by the formula in eq.(24) these input values were used to
calculate the ``experimental" left hand side of eq.(24) for $n=8$, 12,
16 and 20. The choice of these particular values is determined,
in addition to the discussed restrictions on the range of $n$, by the
requirement that the chosen moments are sufficiently representative of
the statistically independent input data. In other words, if the values
of $n$ were chosen in a shorter range with smaller spacing they would
essentially represent less statistically independent values than there
are experimental inputs. Quantitatively this choice is determined by the
condition number of the covariance matrix of the values of the moments
with respect to the dispersion of inputs.

The actual fit was done in several different ways as regards handling of
the uncertainty due to the $O(1/n)$ terms in eq.(\ref{rf}) and the
uncertainty in the continuum cross section. The effect of the former
terms was parametrized as a factor $(1+c/n)$ in the right hand side of
eq.(\ref{rf}) and the value of the coefficient $c$ was either treated as
a fit parameter, or fixed at $c=0$. Then the difference in the fit values
of $\as$ and $m_b$ in these two calculations gives the estimate of the
uncertainty due to the unknown $O(1/n)$ terms. The 50\% uncertainty in
the cross section in the continuum was treated either as an additional
statistical error, or the extreme cases were taken as fixed.
For the $\as$ all fits were done in terms of $\ams(1\, GeV)$.

The results of the fit, where the uncertainty in the continuum cross
section is treated as a statistical error are the following. A three
parameter fit for $\as$, $m_b$ and $c$ yields: $\ams(1\, GeV)=0.336 \pm
0.006$, $m_b=4827 \pm 4 \, MeV$ and $c=-0.59 \pm 0.19$ with $\chi^2=0.8$
at the minimum corresponding to the central values of the fit
parameters. If $c$ is fixed at $c=0$ and a two parameter fit is
performed, the minimum of $\chi2$ is reached at $\ams(1\, GeV)=0.325$
and $m_b=4280 \, MeV$. However the value of $\chi^2$ at the minimum is
$\chi^2=7.2$, thus it would not be reasonable to ascribe statistical
errors to the fit parameters in this case. However, the difference
of the central values for $\as$ and $m_b$ in the two fits
gives the measure of the uncertainty due to
the $O(1/n)$ term, which thus leads to the estimate of the errors:
\begin{eqnarray}
&& \ams(1\, GeV)=0.336 \pm 0.011 \nonumber \\
&& m_b=4827 \pm 7 \, MeV~.
\label{fina}
\end{eqnarray}
A plot, illustrating these two fits, is shown in Fig. 4.

If the continuum cross section is fixed at the lower extreme bound:
$R_b(s)=0.5/3$ at $\sqrt{s} > M(4S)$, the result of a three parameter
fit is $\ams(1\, GeV)=0.338 \pm 0.005$, $m_b=4828 \pm 4 \, MeV$,
$c=-0.71 \pm 0.16$ with $\chi^2=0.1$ at the minimum, which values are
within the range given by eq.(\ref{fina}). However, if $c$ is fixed at
$c=0$ the two parameter fit gives a statistically unacceptable minimal
value $\chi^2=424$ at $\ams(1\, GeV)=0.313$ and $m_b=4809 \, MeV$.
Setting the possible continuum cross section at the upper extreme value:
$R_b(s)=1.5/3$ does not give at all a statistically acceptable fit
for either
three or two parameters. (In the former case one finds $\chi^2=43$ at the
minimum at $\ams(1\, GeV)=0.307$ and $m_b=4808 MeV$, while a two
parameter fit gives a totally unreasonable
minimal value $\chi^2=800$.) As a result
we conclude that with all the uncertainty in the continuum cross section
and in the $O(1/n)$ terms in the sum rules (\ref{rf}) taken into account
the existing data on $\Upsilon$ resonances are only compatible with
values of $\as$ and $m_b$, which lie in the range given by
eq.(\ref{fina}).

\section{Discussion}

In the derivation of the sum rules in eq.(\ref{rf}) only a one-loop
running of $\as$ is taken into account. Therefore
it might be possibly argued that the result for $\as$ found in the present
analysis is not appropriate for a precision two- or three-loop evolution
to higher scales, in particular for comparison with other estimates at
the $Z$ mass scale. To answer to this argument it is instructive to
evaluate the actual interval of the normalization scale over which the
$\as$ evolves in eq.(\ref{rf}) when $n$ changes in the considered range,
i.e. from $n=8$ to $n=20$. Using the central fit values for
$\as$ and $m_b$ in eq.(\ref{fina}),
we find that the normalization point $\mu(n)= m_b \, h(\gamma)
/\sqrt{n}$ changes from $\mu(8)=1.028 \, GeV$ down to $\mu(20)=0.99 \,
GeV$ (for reference: the corresponding values of the Coulomb parameter
$\gamma$ are $\gamma=0.625$ for $n=8$ and $\gamma=1.007$ at $n=20$).
Therefore change in $\ln \mu$ is only about 0.04, so that higher loop
effects in the $\beta$ function would be well less than the error for
$\as$ in eq.(\ref{fina}). Naturally, a comparison
of the present result with other estimates
of $\as$ at different scales involves evolution of the coupling at
large intervals of the normalization scale and should be done using the
higher loop effects. Considering the evolution in two loops with the
starting value of $\ams(1 \, GeV)$ in eq.(\ref{rf}), one finds:
\begin{eqnarray}
&& \Lambda_{\overline {MS}}^{(3)}=258 \pm 14 \, MeV~, \nonumber \\
&& \ams(m_\tau)=0.254 \pm 0.006 ~, \nonumber \\
&& \ams(m_b)=0.185 \pm 0.003 ~, \nonumber \\
&& \ams(M_Z)=0.109 \pm 0.001 ~,
\label{rp}
\end{eqnarray}
where $\Lambda_{\overline {MS}}$ is defined as the position of the
singularity in formal numerical solution of the two-loop
evolution equation for $\as$. These derived values of $\as$ are
significantly lower than the central values of those found from the $\tau$
decay$^{\cite{pich}}$: $\ams(m_\tau)=0.33 \pm 0.03$, and from the LEP
data$^{\cite{vys}}$: $\ams(M_Z)=0.125 \pm 0.005 \pm 0.002$,
although there is a compatibility at the level of about $3\sigma$ due to
larger errors of the latter determinations. On the other hand the value
of $\as$ is in a very good agreement with the one
determined from the three-gluon decay rates of
the $\Upsilon$ resonances, which corresponds to $\ams(M_Z)=0.108 \pm
0.001$ (statistical error only) and which is often doubted as being
subject to unknown non-perturbative and relativistic corrections, as is
discussed in \cite{hinchliffe1}. I believe that the approach
used in this paper is intrinsically more accurate, than determining the
value of $\as$ from small corrections. The reliability of the considered
method can be further improved by experimentally measuring the cross
section of $e^+\, e^-$ annihilation above the $\Upsilon(4S)$ up to about
12 GeV c.m. energy and by a theoretical analysis of the $1/n$ and $\as^2$
terms.

The value of the quark mass in eq.(\ref{fina}) requires some
specification, in view of the intrinsic uncertainty of the quark mass in
non-perturbative QCD$^{\cite{pole}}$. Apart from summation of the Coulomb
terms, the sum rules in eq.(\ref{rf}) contain the QCD radiative effects
of the first order in $\as$. The on shell mass of a heavy quark is
defined in any finite order of perturbation theory and enter as a formal
parameter in a calculation to that order. In this sense the mass
parameter $m_b$ is the on shell mass appropriate for calculations at the
one-loop radiative level in QCD. Naturally, the derived on shell mass in
perturbation theory depends on the order in $\as$ in which the
calculation is done. This is because a perturbative calculation is
justified inasmuch as it is determined by the short-distance dynamics
and thus it is in fact sensitive to an off-shell value of the quark mass
$m(\mu)$ at short distances, which is a combination of the on-shell mass,
the coupling constant $\as(\mu)$
and the momentum scale $\mu$ corresponding to the
distance scale involved in the problem. In particular,
for a heavy quark and $\mu \ll m$ the first-order relation is
$m(\mu)=m-a\, \as(\mu) \, \mu$~, where the constant $a$ is scheme
dependent as the off-shell mass $m(\mu)$ is, unlike the on-shell mass
$m$. Therefore the on-shell mass
derived from a short-distance calculation should correlate with the
value of the coupling $\as(\mu)$.
This correlation is conspicuously present in
the considered here analysis of the sum rules (\ref{rf}), and can be
clearly seen on the plot of Fig. 4. Within the described
numerical analysis, one can
find that an uncorrelated with $\as$ mass parameter is $m_b^*=m_b - 0.56
\, \ams(\mu) \, \mu$ for $\mu = 1 \, GeV$, which is determined with a
very high statistical accuracy: $m_b^*=4639 \pm 2 \, MeV$, which is due
to the fact that the sum rules in the considered range of $n$ are
sensitive to dynamics at distances approximately 1 GeV$^{-1}$. In view
of the scheme dependence of the off-shell mass the precise implications
of this numerical observation are not quite clear.\footnote{It is however
interesting to note, in this connection, that the numerical
value 0.56 of the coefficient $a$ is very close to the value
$a=16/(9 \pi) \approx 0.567$, which enters the definition of the
off-shell mass by the amplitude of Thomson scattering$^{\cite{v4}}$.}

\section{Acknowledgements}
I am thankful to L.B. Okun, M.A. Shifman, M.I. Vysotsky and N.G.
Uraltsev for numerous stimulating discussions.
This work is supported, in part, by the DOE grant DE-AC02-83ER40105.

{\Large \bf Figure Captions}\\[0.2in]

\noindent
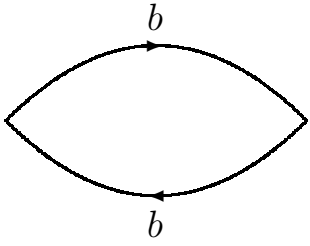
{\bf Figure 1.} The lowest order graph for vacuum polarization by
the vector current of $b$ quarks.\\[0.1in]

\noindent
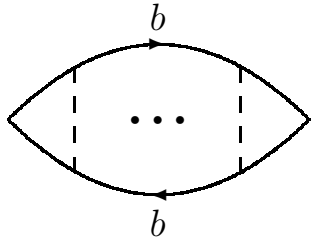
{\bf Figure 2.} The type of graphs corresponding to summation of
all powers of the Coulomb parameter $\as/v$ in the vacuum polarisation
in the non-relativistic domain near the threshold. Dashed lines denote
Coulomb gluons.\\[0.1in]

\noindent
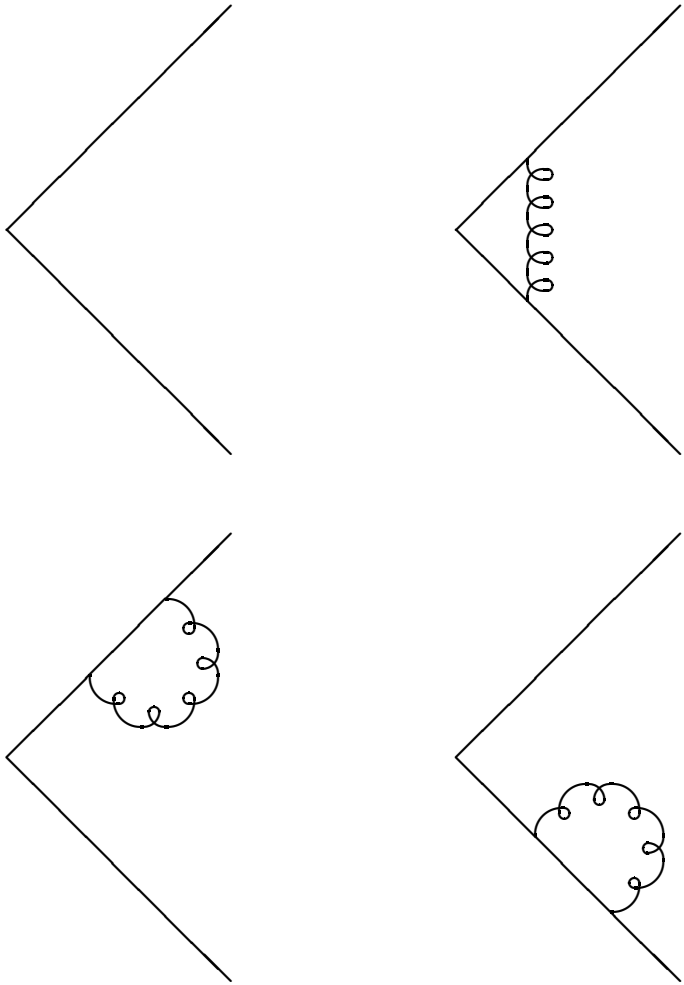
{\bf Figure 3.} The graphs, describing up to the order $\as$ the matrix
element for production of quark and antiquark by the vector current
exactly at the threshold.\\[0.1in]

\noindent
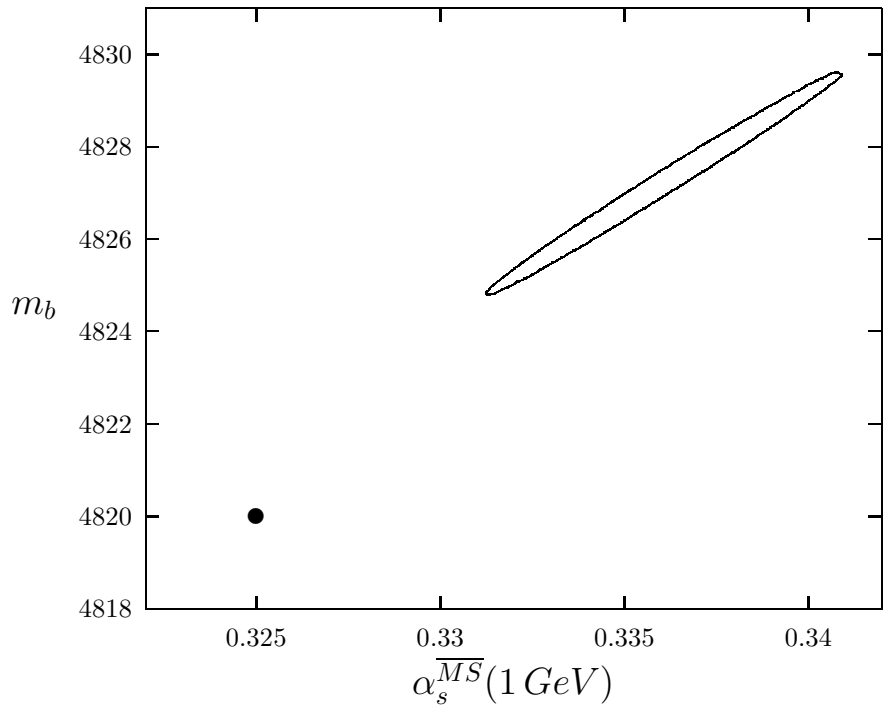
{\bf Figure 4.} The contour plot illustrating the fit of the sum rules
to the data. (The uncertainty in the continuum cross section above the
threshold is treated as a statistical error.) The ellipse is the one
standard deviation contour for $\ams$ and $m_b$ (in MeV)
for the three parameter fit at the optimal
value of $c$: $c=-0.59$. The heavy dot corresponds to the minimum of
$\chi^2$ in the two parameter fit.

\newpage
\unitlength=1mm
\thicklines
\begin{picture}(90.00,110.00)
\bezier{228}(50.00,90.00)(70.00,110.00)(90.00,90.00)
\bezier{228}(50.00,90.00)(70.00,70.00)(90.00,90.00)
\put(69.00,100.00){\vector(1,0){2.00}}
\put(71.00,80.00){\vector(-1,0){2.00}}
\put(70.00,102.00){\makebox(0,0)[cb]{{\large $b$}}}
\put(70.00,78.00){\makebox(0,0)[ct]{{\large $b$}}}
\put(70.00,44.00){\makebox(0,0)[cc]{{\Large \bf Fig. 1}}}
\end{picture}\\[-0.5 in]
\begin{picture}(90.00,110.00)
\bezier{228}(50.00,90.00)(70.00,110.00)(90.00,90.00)
\bezier{228}(50.00,90.00)(70.00,70.00)(90.00,90.00)
\put(69.00,100.00){\vector(1,0){2.00}}
\put(71.00,80.00){\vector(-1,0){2.00}}
\put(70.00,102.00){\makebox(0,0)[cb]{{\large $b$}}}
\put(70.00,78.00){\makebox(0,0)[ct]{{\large $b$}}}
\put(70.00,44.00){\makebox(0,0)[cc]{{\Large \bf Fig. 2}}}
\multiput(59.00,97.00)(0.0,-4.0){4}{\line(0,-1){2.00}}
\multiput(81.00,97.00)(0.0,-4.0){4}{\line(0,-1){2.00}}
\multiput(67.00,90.00)(3.00,0.00){3}{\circle*{1.00}}
\end{picture}
\newpage
\input{FEYNMAN}
\begin{picture}(30000,10000)(0,3000)
\THICKLINES
\drawline\fermion[\NE\REG](3000,5000)[12000]
\drawline\fermion[\SE\REG](3000,5000)[12000]
\drawline\fermion[\NE\REG](20000,5000)[12000]
\drawline\fermion[\SE\REG](20000,5000)[12000]
\drawline\gluon[\S\FLIPPED](22700,7700)[5]
\drawline\fermion[\NE\REG](3000,-15000)[12000]
\drawline\fermion[\SE\REG](3000,-15000)[12000]
\drawloop\gluon[\SE 5](9000,-9000)
\drawline\fermion[\NE\REG](20000,-15000)[12000]
\drawline\fermion[\SE\REG](20000,-15000)[12000]
\drawloop\gluon[\NE 5](23000,-18000)
\put(15000,-35000){\Large\bf Fig. 3}
\end{picture}
\newpage
%
\setlength{\unitlength}{0.240900pt}
\ifx\plotpoint\undefined\newsavebox{\plotpoint}\fi
\begin{picture}(1439,1080)(0,0)
\font\gnuplot=cmr10 at 10pt
\gnuplot
\sbox{\plotpoint}{\rule[-0.200pt]{0.400pt}{0.400pt}}%
\put(220.0,113.0){\rule[-0.200pt]{4.818pt}{0.400pt}}
\put(198,113){\makebox(0,0)[r]{4818}}
\put(1355.0,113.0){\rule[-0.200pt]{4.818pt}{0.400pt}}
\put(220.0,258.0){\rule[-0.200pt]{4.818pt}{0.400pt}}
\put(198,258){\makebox(0,0)[r]{4820}}
\put(1355.0,258.0){\rule[-0.200pt]{4.818pt}{0.400pt}}
\put(220.0,403.0){\rule[-0.200pt]{4.818pt}{0.400pt}}
\put(198,403){\makebox(0,0)[r]{4822}}
\put(1355.0,403.0){\rule[-0.200pt]{4.818pt}{0.400pt}}
\put(220.0,549.0){\rule[-0.200pt]{4.818pt}{0.400pt}}
\put(198,549){\makebox(0,0)[r]{4824}}
\put(1355.0,549.0){\rule[-0.200pt]{4.818pt}{0.400pt}}
\put(220.0,694.0){\rule[-0.200pt]{4.818pt}{0.400pt}}
\put(198,694){\makebox(0,0)[r]{4826}}
\put(1355.0,694.0){\rule[-0.200pt]{4.818pt}{0.400pt}}
\put(220.0,839.0){\rule[-0.200pt]{4.818pt}{0.400pt}}
\put(198,839){\makebox(0,0)[r]{4828}}
\put(1355.0,839.0){\rule[-0.200pt]{4.818pt}{0.400pt}}
\put(220.0,984.0){\rule[-0.200pt]{4.818pt}{0.400pt}}
\put(198,984){\makebox(0,0)[r]{4830}}
\put(1355.0,984.0){\rule[-0.200pt]{4.818pt}{0.400pt}}
\put(393.0,113.0){\rule[-0.200pt]{0.400pt}{4.818pt}}
\put(393,68){\makebox(0,0){0.325}}
\put(393.0,1037.0){\rule[-0.200pt]{0.400pt}{4.818pt}}
\put(682.0,113.0){\rule[-0.200pt]{0.400pt}{4.818pt}}
\put(682,68){\makebox(0,0){0.33}}
\put(682.0,1037.0){\rule[-0.200pt]{0.400pt}{4.818pt}}
\put(971.0,113.0){\rule[-0.200pt]{0.400pt}{4.818pt}}
\put(971,68){\makebox(0,0){0.335}}
\put(971.0,1037.0){\rule[-0.200pt]{0.400pt}{4.818pt}}
\put(1260.0,113.0){\rule[-0.200pt]{0.400pt}{4.818pt}}
\put(1260,68){\makebox(0,0){0.34}}
\put(1260.0,1037.0){\rule[-0.200pt]{0.400pt}{4.818pt}}
\put(220.0,113.0){\rule[-0.200pt]{278.239pt}{0.400pt}}
\put(1375.0,113.0){\rule[-0.200pt]{0.400pt}{227.410pt}}
\put(220.0,1057.0){\rule[-0.200pt]{278.239pt}{0.400pt}}
\put(45,585){\makebox(0,0){{\large $m_b$}}}
\put(797,0.0){\makebox(0,0){{\large $\ams(1\, GeV)$}}}
\put(220.0,113.0){\rule[-0.200pt]{0.400pt}{227.410pt}}
\put(1305,956){\usebox{\plotpoint}}
\multiput(1297.94,954.95)(-2.472,-0.447){3}{\rule{1.700pt}{0.108pt}}
\multiput(1301.47,955.17)(-8.472,-3.000){2}{\rule{0.850pt}{0.400pt}}
\put(1292,951.67){\rule{0.241pt}{0.400pt}}
\multiput(1292.50,952.17)(-0.500,-1.000){2}{\rule{0.120pt}{0.400pt}}
\multiput(1288.82,950.93)(-0.852,-0.482){9}{\rule{0.767pt}{0.116pt}}
\multiput(1290.41,951.17)(-8.409,-6.000){2}{\rule{0.383pt}{0.400pt}}
\multiput(1276.60,944.95)(-1.802,-0.447){3}{\rule{1.300pt}{0.108pt}}
\multiput(1279.30,945.17)(-6.302,-3.000){2}{\rule{0.650pt}{0.400pt}}
\put(1270,941.17){\rule{0.700pt}{0.400pt}}
\multiput(1271.55,942.17)(-1.547,-2.000){2}{\rule{0.350pt}{0.400pt}}
\multiput(1266.26,939.93)(-1.033,-0.482){9}{\rule{0.900pt}{0.116pt}}
\multiput(1268.13,940.17)(-10.132,-6.000){2}{\rule{0.450pt}{0.400pt}}
\put(1255,933.17){\rule{0.700pt}{0.400pt}}
\multiput(1256.55,934.17)(-1.547,-2.000){2}{\rule{0.350pt}{0.400pt}}
\multiput(1251.93,931.93)(-0.821,-0.477){7}{\rule{0.740pt}{0.115pt}}
\multiput(1253.46,932.17)(-6.464,-5.000){2}{\rule{0.370pt}{0.400pt}}
\multiput(1242.43,926.94)(-1.358,-0.468){5}{\rule{1.100pt}{0.113pt}}
\multiput(1244.72,927.17)(-7.717,-4.000){2}{\rule{0.550pt}{0.400pt}}
\put(1235,922.17){\rule{0.482pt}{0.400pt}}
\multiput(1236.00,923.17)(-1.000,-2.000){2}{\rule{0.241pt}{0.400pt}}
\multiput(1231.26,920.93)(-1.033,-0.482){9}{\rule{0.900pt}{0.116pt}}
\multiput(1233.13,921.17)(-10.132,-6.000){2}{\rule{0.450pt}{0.400pt}}
\put(1220,914.17){\rule{0.700pt}{0.400pt}}
\multiput(1221.55,915.17)(-1.547,-2.000){2}{\rule{0.350pt}{0.400pt}}
\multiput(1216.93,912.93)(-0.821,-0.477){7}{\rule{0.740pt}{0.115pt}}
\multiput(1218.46,913.17)(-6.464,-5.000){2}{\rule{0.370pt}{0.400pt}}
\multiput(1208.60,907.93)(-0.933,-0.477){7}{\rule{0.820pt}{0.115pt}}
\multiput(1210.30,908.17)(-7.298,-5.000){2}{\rule{0.410pt}{0.400pt}}
\put(1200,902.67){\rule{0.723pt}{0.400pt}}
\multiput(1201.50,903.17)(-1.500,-1.000){2}{\rule{0.361pt}{0.400pt}}
\multiput(1196.74,901.93)(-0.874,-0.485){11}{\rule{0.786pt}{0.117pt}}
\multiput(1198.37,902.17)(-10.369,-7.000){2}{\rule{0.393pt}{0.400pt}}
\put(1186,894.67){\rule{0.482pt}{0.400pt}}
\multiput(1187.00,895.17)(-1.000,-1.000){2}{\rule{0.241pt}{0.400pt}}
\multiput(1183.09,893.93)(-0.762,-0.482){9}{\rule{0.700pt}{0.116pt}}
\multiput(1184.55,894.17)(-7.547,-6.000){2}{\rule{0.350pt}{0.400pt}}
\multiput(1173.26,887.94)(-1.066,-0.468){5}{\rule{0.900pt}{0.113pt}}
\multiput(1175.13,888.17)(-6.132,-4.000){2}{\rule{0.450pt}{0.400pt}}
\put(1165,883.17){\rule{0.900pt}{0.400pt}}
\multiput(1167.13,884.17)(-2.132,-2.000){2}{\rule{0.450pt}{0.400pt}}
\multiput(1161.74,881.93)(-0.874,-0.485){11}{\rule{0.786pt}{0.117pt}}
\multiput(1163.37,882.17)(-10.369,-7.000){2}{\rule{0.393pt}{0.400pt}}
\put(1153,876){\usebox{\plotpoint}}
\multiput(1149.98,874.93)(-0.798,-0.485){11}{\rule{0.729pt}{0.117pt}}
\multiput(1151.49,875.17)(-9.488,-7.000){2}{\rule{0.364pt}{0.400pt}}
\multiput(1138.82,867.95)(-0.909,-0.447){3}{\rule{0.767pt}{0.108pt}}
\multiput(1140.41,868.17)(-3.409,-3.000){2}{\rule{0.383pt}{0.400pt}}
\multiput(1133.68,864.94)(-0.920,-0.468){5}{\rule{0.800pt}{0.113pt}}
\multiput(1135.34,865.17)(-5.340,-4.000){2}{\rule{0.400pt}{0.400pt}}
\multiput(1126.60,860.93)(-0.933,-0.477){7}{\rule{0.820pt}{0.115pt}}
\multiput(1128.30,861.17)(-7.298,-5.000){2}{\rule{0.410pt}{0.400pt}}
\put(1118,855.17){\rule{0.700pt}{0.400pt}}
\multiput(1119.55,856.17)(-1.547,-2.000){2}{\rule{0.350pt}{0.400pt}}
\multiput(1114.54,853.93)(-0.943,-0.482){9}{\rule{0.833pt}{0.116pt}}
\multiput(1116.27,854.17)(-9.270,-6.000){2}{\rule{0.417pt}{0.400pt}}
\put(1104,847.17){\rule{0.700pt}{0.400pt}}
\multiput(1105.55,848.17)(-1.547,-2.000){2}{\rule{0.350pt}{0.400pt}}
\multiput(1100.60,845.93)(-0.933,-0.477){7}{\rule{0.820pt}{0.115pt}}
\multiput(1102.30,846.17)(-7.298,-5.000){2}{\rule{0.410pt}{0.400pt}}
\multiput(1091.68,840.94)(-0.920,-0.468){5}{\rule{0.800pt}{0.113pt}}
\multiput(1093.34,841.17)(-5.340,-4.000){2}{\rule{0.400pt}{0.400pt}}
\multiput(1085.51,836.94)(-0.627,-0.468){5}{\rule{0.600pt}{0.113pt}}
\multiput(1086.75,837.17)(-3.755,-4.000){2}{\rule{0.300pt}{0.400pt}}
\multiput(1079.54,832.93)(-0.943,-0.482){9}{\rule{0.833pt}{0.116pt}}
\multiput(1081.27,833.17)(-9.270,-6.000){2}{\rule{0.417pt}{0.400pt}}
\put(1072,828){\usebox{\plotpoint}}
\multiput(1069.09,826.93)(-0.758,-0.488){13}{\rule{0.700pt}{0.117pt}}
\multiput(1070.55,827.17)(-10.547,-8.000){2}{\rule{0.350pt}{0.400pt}}
\put(1057,818.67){\rule{0.723pt}{0.400pt}}
\multiput(1058.50,819.17)(-1.500,-1.000){2}{\rule{0.361pt}{0.400pt}}
\multiput(1054.09,817.93)(-0.762,-0.482){9}{\rule{0.700pt}{0.116pt}}
\multiput(1055.55,818.17)(-7.547,-6.000){2}{\rule{0.350pt}{0.400pt}}
\multiput(1045.09,811.94)(-0.774,-0.468){5}{\rule{0.700pt}{0.113pt}}
\multiput(1046.55,812.17)(-4.547,-4.000){2}{\rule{0.350pt}{0.400pt}}
\multiput(1038.82,807.95)(-0.909,-0.447){3}{\rule{0.767pt}{0.108pt}}
\multiput(1040.41,808.17)(-3.409,-3.000){2}{\rule{0.383pt}{0.400pt}}
\multiput(1033.54,804.93)(-0.943,-0.482){9}{\rule{0.833pt}{0.116pt}}
\multiput(1035.27,805.17)(-9.270,-6.000){2}{\rule{0.417pt}{0.400pt}}
\put(1025,798.67){\rule{0.241pt}{0.400pt}}
\multiput(1025.50,799.17)(-0.500,-1.000){2}{\rule{0.120pt}{0.400pt}}
\multiput(1022.09,797.93)(-0.758,-0.488){13}{\rule{0.700pt}{0.117pt}}
\multiput(1023.55,798.17)(-10.547,-8.000){2}{\rule{0.350pt}{0.400pt}}
\put(1011,789.67){\rule{0.482pt}{0.400pt}}
\multiput(1012.00,790.17)(-1.000,-1.000){2}{\rule{0.241pt}{0.400pt}}
\multiput(1008.09,788.93)(-0.762,-0.482){9}{\rule{0.700pt}{0.116pt}}
\multiput(1009.55,789.17)(-7.547,-6.000){2}{\rule{0.350pt}{0.400pt}}
\multiput(999.09,782.94)(-0.774,-0.468){5}{\rule{0.700pt}{0.113pt}}
\multiput(1000.55,783.17)(-4.547,-4.000){2}{\rule{0.350pt}{0.400pt}}
\multiput(993.09,778.94)(-0.774,-0.468){5}{\rule{0.700pt}{0.113pt}}
\multiput(994.55,779.17)(-4.547,-4.000){2}{\rule{0.350pt}{0.400pt}}
\multiput(986.60,774.93)(-0.933,-0.477){7}{\rule{0.820pt}{0.115pt}}
\multiput(988.30,775.17)(-7.298,-5.000){2}{\rule{0.410pt}{0.400pt}}
\put(978,769.17){\rule{0.700pt}{0.400pt}}
\multiput(979.55,770.17)(-1.547,-2.000){2}{\rule{0.350pt}{0.400pt}}
\multiput(974.98,767.93)(-0.798,-0.485){11}{\rule{0.729pt}{0.117pt}}
\multiput(976.49,768.17)(-9.488,-7.000){2}{\rule{0.364pt}{0.400pt}}
\put(966,760.67){\rule{0.241pt}{0.400pt}}
\multiput(966.50,761.17)(-0.500,-1.000){2}{\rule{0.120pt}{0.400pt}}
\multiput(962.98,759.93)(-0.798,-0.485){11}{\rule{0.729pt}{0.117pt}}
\multiput(964.49,760.17)(-9.488,-7.000){2}{\rule{0.364pt}{0.400pt}}
\put(952,752.17){\rule{0.700pt}{0.400pt}}
\multiput(953.55,753.17)(-1.547,-2.000){2}{\rule{0.350pt}{0.400pt}}
\multiput(949.09,750.93)(-0.762,-0.482){9}{\rule{0.700pt}{0.116pt}}
\multiput(950.55,751.17)(-7.547,-6.000){2}{\rule{0.350pt}{0.400pt}}
\multiput(940.09,744.94)(-0.774,-0.468){5}{\rule{0.700pt}{0.113pt}}
\multiput(941.55,745.17)(-4.547,-4.000){2}{\rule{0.350pt}{0.400pt}}
\multiput(934.51,740.94)(-0.627,-0.468){5}{\rule{0.600pt}{0.113pt}}
\multiput(935.75,741.17)(-3.755,-4.000){2}{\rule{0.300pt}{0.400pt}}
\multiput(928.26,736.93)(-1.044,-0.477){7}{\rule{0.900pt}{0.115pt}}
\multiput(930.13,737.17)(-8.132,-5.000){2}{\rule{0.450pt}{0.400pt}}
\put(920,731.17){\rule{0.482pt}{0.400pt}}
\multiput(921.00,732.17)(-1.000,-2.000){2}{\rule{0.241pt}{0.400pt}}
\multiput(916.74,729.93)(-0.874,-0.485){11}{\rule{0.786pt}{0.117pt}}
\multiput(918.37,730.17)(-10.369,-7.000){2}{\rule{0.393pt}{0.400pt}}
\multiput(904.98,721.93)(-0.798,-0.485){11}{\rule{0.729pt}{0.117pt}}
\multiput(906.49,722.17)(-9.488,-7.000){2}{\rule{0.364pt}{0.400pt}}
\put(894,714.17){\rule{0.700pt}{0.400pt}}
\multiput(895.55,715.17)(-1.547,-2.000){2}{\rule{0.350pt}{0.400pt}}
\multiput(891.09,712.93)(-0.762,-0.482){9}{\rule{0.700pt}{0.116pt}}
\multiput(892.55,713.17)(-7.547,-6.000){2}{\rule{0.350pt}{0.400pt}}
\multiput(882.51,706.94)(-0.627,-0.468){5}{\rule{0.600pt}{0.113pt}}
\multiput(883.75,707.17)(-3.755,-4.000){2}{\rule{0.300pt}{0.400pt}}
\multiput(876.68,702.94)(-0.920,-0.468){5}{\rule{0.800pt}{0.113pt}}
\multiput(878.34,703.17)(-5.340,-4.000){2}{\rule{0.400pt}{0.400pt}}
\multiput(870.26,698.93)(-0.710,-0.477){7}{\rule{0.660pt}{0.115pt}}
\multiput(871.63,699.17)(-5.630,-5.000){2}{\rule{0.330pt}{0.400pt}}
\multiput(863.37,693.95)(-0.685,-0.447){3}{\rule{0.633pt}{0.108pt}}
\multiput(864.69,694.17)(-2.685,-3.000){2}{\rule{0.317pt}{0.400pt}}
\multiput(858.98,690.93)(-0.798,-0.485){11}{\rule{0.729pt}{0.117pt}}
\multiput(860.49,691.17)(-9.488,-7.000){2}{\rule{0.364pt}{0.400pt}}
\put(850,683.67){\rule{0.241pt}{0.400pt}}
\multiput(850.50,684.17)(-0.500,-1.000){2}{\rule{0.120pt}{0.400pt}}
\multiput(847.09,682.93)(-0.758,-0.488){13}{\rule{0.700pt}{0.117pt}}
\multiput(848.55,683.17)(-10.547,-8.000){2}{\rule{0.350pt}{0.400pt}}
\put(908.0,723.0){\usebox{\plotpoint}}
\put(838,676){\usebox{\plotpoint}}
\multiput(835.30,674.93)(-0.692,-0.488){13}{\rule{0.650pt}{0.117pt}}
\multiput(836.65,675.17)(-9.651,-8.000){2}{\rule{0.325pt}{0.400pt}}
\put(824,666.17){\rule{0.700pt}{0.400pt}}
\multiput(825.55,667.17)(-1.547,-2.000){2}{\rule{0.350pt}{0.400pt}}
\multiput(821.45,664.93)(-0.645,-0.485){11}{\rule{0.614pt}{0.117pt}}
\multiput(822.73,665.17)(-7.725,-7.000){2}{\rule{0.307pt}{0.400pt}}
\put(811,657.17){\rule{0.900pt}{0.400pt}}
\multiput(813.13,658.17)(-2.132,-2.000){2}{\rule{0.450pt}{0.400pt}}
\multiput(808.69,655.93)(-0.569,-0.485){11}{\rule{0.557pt}{0.117pt}}
\multiput(809.84,656.17)(-6.844,-7.000){2}{\rule{0.279pt}{0.400pt}}
\multiput(799.82,648.95)(-0.909,-0.447){3}{\rule{0.767pt}{0.108pt}}
\multiput(801.41,649.17)(-3.409,-3.000){2}{\rule{0.383pt}{0.400pt}}
\multiput(795.59,645.93)(-0.599,-0.477){7}{\rule{0.580pt}{0.115pt}}
\multiput(796.80,646.17)(-4.796,-5.000){2}{\rule{0.290pt}{0.400pt}}
\multiput(789.59,640.93)(-0.599,-0.477){7}{\rule{0.580pt}{0.115pt}}
\multiput(790.80,641.17)(-4.796,-5.000){2}{\rule{0.290pt}{0.400pt}}
\multiput(783.09,635.94)(-0.774,-0.468){5}{\rule{0.700pt}{0.113pt}}
\multiput(784.55,636.17)(-4.547,-4.000){2}{\rule{0.350pt}{0.400pt}}
\multiput(777.59,631.93)(-0.599,-0.477){7}{\rule{0.580pt}{0.115pt}}
\multiput(778.80,632.17)(-4.796,-5.000){2}{\rule{0.290pt}{0.400pt}}
\multiput(771.59,626.93)(-0.599,-0.477){7}{\rule{0.580pt}{0.115pt}}
\multiput(772.80,627.17)(-4.796,-5.000){2}{\rule{0.290pt}{0.400pt}}
\multiput(765.59,621.93)(-0.599,-0.477){7}{\rule{0.580pt}{0.115pt}}
\multiput(766.80,622.17)(-4.796,-5.000){2}{\rule{0.290pt}{0.400pt}}
\multiput(760.93,615.59)(-0.477,-0.599){7}{\rule{0.115pt}{0.580pt}}
\multiput(761.17,616.80)(-5.000,-4.796){2}{\rule{0.400pt}{0.290pt}}
\multiput(754.37,610.95)(-0.685,-0.447){3}{\rule{0.633pt}{0.108pt}}
\multiput(755.69,611.17)(-2.685,-3.000){2}{\rule{0.317pt}{0.400pt}}
\multiput(753.00,607.95)(0.685,-0.447){3}{\rule{0.633pt}{0.108pt}}
\multiput(753.00,608.17)(2.685,-3.000){2}{\rule{0.317pt}{0.400pt}}
\put(757,606.17){\rule{2.300pt}{0.400pt}}
\multiput(757.00,605.17)(6.226,2.000){2}{\rule{1.150pt}{0.400pt}}
\put(768,607.67){\rule{0.241pt}{0.400pt}}
\multiput(768.00,607.17)(0.500,1.000){2}{\rule{0.120pt}{0.400pt}}
\multiput(769.00,609.59)(0.943,0.482){9}{\rule{0.833pt}{0.116pt}}
\multiput(769.00,608.17)(9.270,6.000){2}{\rule{0.417pt}{0.400pt}}
\multiput(780.00,615.61)(1.802,0.447){3}{\rule{1.300pt}{0.108pt}}
\multiput(780.00,614.17)(6.302,3.000){2}{\rule{0.650pt}{0.400pt}}
\put(789,618.17){\rule{0.700pt}{0.400pt}}
\multiput(789.00,617.17)(1.547,2.000){2}{\rule{0.350pt}{0.400pt}}
\multiput(792.00,620.59)(0.943,0.482){9}{\rule{0.833pt}{0.116pt}}
\multiput(792.00,619.17)(9.270,6.000){2}{\rule{0.417pt}{0.400pt}}
\put(803,626.17){\rule{0.900pt}{0.400pt}}
\multiput(803.00,625.17)(2.132,2.000){2}{\rule{0.450pt}{0.400pt}}
\multiput(807.00,628.60)(1.066,0.468){5}{\rule{0.900pt}{0.113pt}}
\multiput(807.00,627.17)(6.132,4.000){2}{\rule{0.450pt}{0.400pt}}
\multiput(815.00,632.59)(1.155,0.477){7}{\rule{0.980pt}{0.115pt}}
\multiput(815.00,631.17)(8.966,5.000){2}{\rule{0.490pt}{0.400pt}}
\put(826,636.67){\rule{0.241pt}{0.400pt}}
\multiput(826.00,636.17)(0.500,1.000){2}{\rule{0.120pt}{0.400pt}}
\multiput(827.00,638.59)(0.798,0.485){11}{\rule{0.729pt}{0.117pt}}
\multiput(827.00,637.17)(9.488,7.000){2}{\rule{0.364pt}{0.400pt}}
\put(838,645.17){\rule{0.900pt}{0.400pt}}
\multiput(838.00,644.17)(2.132,2.000){2}{\rule{0.450pt}{0.400pt}}
\multiput(842.00,647.59)(0.821,0.477){7}{\rule{0.740pt}{0.115pt}}
\multiput(842.00,646.17)(6.464,5.000){2}{\rule{0.370pt}{0.400pt}}
\multiput(850.00,652.59)(0.933,0.477){7}{\rule{0.820pt}{0.115pt}}
\multiput(850.00,651.17)(7.298,5.000){2}{\rule{0.410pt}{0.400pt}}
\put(859,656.67){\rule{0.723pt}{0.400pt}}
\multiput(859.00,656.17)(1.500,1.000){2}{\rule{0.361pt}{0.400pt}}
\multiput(862.00,658.59)(0.798,0.485){11}{\rule{0.729pt}{0.117pt}}
\multiput(862.00,657.17)(9.488,7.000){2}{\rule{0.364pt}{0.400pt}}
\put(873,664.67){\rule{0.723pt}{0.400pt}}
\multiput(873.00,664.17)(1.500,1.000){2}{\rule{0.361pt}{0.400pt}}
\multiput(876.00,666.59)(0.762,0.482){9}{\rule{0.700pt}{0.116pt}}
\multiput(876.00,665.17)(7.547,6.000){2}{\rule{0.350pt}{0.400pt}}
\multiput(885.00,672.60)(0.920,0.468){5}{\rule{0.800pt}{0.113pt}}
\multiput(885.00,671.17)(5.340,4.000){2}{\rule{0.400pt}{0.400pt}}
\put(892,676.17){\rule{1.100pt}{0.400pt}}
\multiput(892.00,675.17)(2.717,2.000){2}{\rule{0.550pt}{0.400pt}}
\multiput(897.00,678.59)(0.798,0.485){11}{\rule{0.729pt}{0.117pt}}
\multiput(897.00,677.17)(9.488,7.000){2}{\rule{0.364pt}{0.400pt}}
\multiput(909.00,685.59)(0.798,0.485){11}{\rule{0.729pt}{0.117pt}}
\multiput(909.00,684.17)(9.488,7.000){2}{\rule{0.364pt}{0.400pt}}
\multiput(920.00,692.61)(0.909,0.447){3}{\rule{0.767pt}{0.108pt}}
\multiput(920.00,691.17)(3.409,3.000){2}{\rule{0.383pt}{0.400pt}}
\multiput(925.00,695.60)(0.920,0.468){5}{\rule{0.800pt}{0.113pt}}
\multiput(925.00,694.17)(5.340,4.000){2}{\rule{0.400pt}{0.400pt}}
\multiput(932.00,699.59)(0.933,0.477){7}{\rule{0.820pt}{0.115pt}}
\multiput(932.00,698.17)(7.298,5.000){2}{\rule{0.410pt}{0.400pt}}
\put(941,704.17){\rule{0.482pt}{0.400pt}}
\multiput(941.00,703.17)(1.000,2.000){2}{\rule{0.241pt}{0.400pt}}
\multiput(943.00,706.59)(0.874,0.485){11}{\rule{0.786pt}{0.117pt}}
\multiput(943.00,705.17)(10.369,7.000){2}{\rule{0.393pt}{0.400pt}}
\put(955,712.67){\rule{0.482pt}{0.400pt}}
\multiput(955.00,712.17)(1.000,1.000){2}{\rule{0.241pt}{0.400pt}}
\multiput(957.00,714.59)(0.852,0.482){9}{\rule{0.767pt}{0.116pt}}
\multiput(957.00,713.17)(8.409,6.000){2}{\rule{0.383pt}{0.400pt}}
\multiput(967.00,720.61)(0.909,0.447){3}{\rule{0.767pt}{0.108pt}}
\multiput(967.00,719.17)(3.409,3.000){2}{\rule{0.383pt}{0.400pt}}
\multiput(972.00,723.60)(0.774,0.468){5}{\rule{0.700pt}{0.113pt}}
\multiput(972.00,722.17)(4.547,4.000){2}{\rule{0.350pt}{0.400pt}}
\multiput(978.00,727.59)(0.762,0.482){9}{\rule{0.700pt}{0.116pt}}
\multiput(978.00,726.17)(7.547,6.000){2}{\rule{0.350pt}{0.400pt}}
\put(987,733.17){\rule{0.700pt}{0.400pt}}
\multiput(987.00,732.17)(1.547,2.000){2}{\rule{0.350pt}{0.400pt}}
\multiput(990.00,735.59)(0.874,0.485){11}{\rule{0.786pt}{0.117pt}}
\multiput(990.00,734.17)(10.369,7.000){2}{\rule{0.393pt}{0.400pt}}
\put(908.0,685.0){\usebox{\plotpoint}}
\multiput(1003.00,742.59)(0.721,0.485){11}{\rule{0.671pt}{0.117pt}}
\multiput(1003.00,741.17)(8.606,7.000){2}{\rule{0.336pt}{0.400pt}}
\multiput(1013.00,749.61)(0.909,0.447){3}{\rule{0.767pt}{0.108pt}}
\multiput(1013.00,748.17)(3.409,3.000){2}{\rule{0.383pt}{0.400pt}}
\multiput(1018.00,752.60)(0.920,0.468){5}{\rule{0.800pt}{0.113pt}}
\multiput(1018.00,751.17)(5.340,4.000){2}{\rule{0.400pt}{0.400pt}}
\multiput(1025.00,756.59)(0.821,0.477){7}{\rule{0.740pt}{0.115pt}}
\multiput(1025.00,755.17)(6.464,5.000){2}{\rule{0.370pt}{0.400pt}}
\multiput(1033.00,761.61)(0.685,0.447){3}{\rule{0.633pt}{0.108pt}}
\multiput(1033.00,760.17)(2.685,3.000){2}{\rule{0.317pt}{0.400pt}}
\multiput(1037.00,764.59)(0.798,0.485){11}{\rule{0.729pt}{0.117pt}}
\multiput(1037.00,763.17)(9.488,7.000){2}{\rule{0.364pt}{0.400pt}}
\put(1002.0,742.0){\usebox{\plotpoint}}
\multiput(1049.00,771.59)(0.798,0.485){11}{\rule{0.729pt}{0.117pt}}
\multiput(1049.00,770.17)(9.488,7.000){2}{\rule{0.364pt}{0.400pt}}
\put(1060,778.17){\rule{0.900pt}{0.400pt}}
\multiput(1060.00,777.17)(2.132,2.000){2}{\rule{0.450pt}{0.400pt}}
\multiput(1064.00,780.59)(0.671,0.482){9}{\rule{0.633pt}{0.116pt}}
\multiput(1064.00,779.17)(6.685,6.000){2}{\rule{0.317pt}{0.400pt}}
\multiput(1072.00,786.60)(0.920,0.468){5}{\rule{0.800pt}{0.113pt}}
\multiput(1072.00,785.17)(5.340,4.000){2}{\rule{0.400pt}{0.400pt}}
\multiput(1079.00,790.61)(0.685,0.447){3}{\rule{0.633pt}{0.108pt}}
\multiput(1079.00,789.17)(2.685,3.000){2}{\rule{0.317pt}{0.400pt}}
\multiput(1083.00,793.59)(0.798,0.485){11}{\rule{0.729pt}{0.117pt}}
\multiput(1083.00,792.17)(9.488,7.000){2}{\rule{0.364pt}{0.400pt}}
\put(1048.0,771.0){\usebox{\plotpoint}}
\multiput(1095.00,800.59)(0.874,0.485){11}{\rule{0.786pt}{0.117pt}}
\multiput(1095.00,799.17)(10.369,7.000){2}{\rule{0.393pt}{0.400pt}}
\put(1107,807.17){\rule{0.482pt}{0.400pt}}
\multiput(1107.00,806.17)(1.000,2.000){2}{\rule{0.241pt}{0.400pt}}
\multiput(1109.00,809.59)(0.762,0.482){9}{\rule{0.700pt}{0.116pt}}
\multiput(1109.00,808.17)(7.547,6.000){2}{\rule{0.350pt}{0.400pt}}
\multiput(1118.00,815.60)(0.774,0.468){5}{\rule{0.700pt}{0.113pt}}
\multiput(1118.00,814.17)(4.547,4.000){2}{\rule{0.350pt}{0.400pt}}
\multiput(1124.00,819.60)(0.774,0.468){5}{\rule{0.700pt}{0.113pt}}
\multiput(1124.00,818.17)(4.547,4.000){2}{\rule{0.350pt}{0.400pt}}
\multiput(1130.00,823.59)(0.933,0.477){7}{\rule{0.820pt}{0.115pt}}
\multiput(1130.00,822.17)(7.298,5.000){2}{\rule{0.410pt}{0.400pt}}
\put(1139,828.17){\rule{0.700pt}{0.400pt}}
\multiput(1139.00,827.17)(1.547,2.000){2}{\rule{0.350pt}{0.400pt}}
\multiput(1142.00,830.59)(0.798,0.485){11}{\rule{0.729pt}{0.117pt}}
\multiput(1142.00,829.17)(9.488,7.000){2}{\rule{0.364pt}{0.400pt}}
\put(1153,836.67){\rule{0.241pt}{0.400pt}}
\multiput(1153.00,836.17)(0.500,1.000){2}{\rule{0.120pt}{0.400pt}}
\multiput(1154.00,838.59)(0.798,0.485){11}{\rule{0.729pt}{0.117pt}}
\multiput(1154.00,837.17)(9.488,7.000){2}{\rule{0.364pt}{0.400pt}}
\put(1165,845.17){\rule{0.700pt}{0.400pt}}
\multiput(1165.00,844.17)(1.547,2.000){2}{\rule{0.350pt}{0.400pt}}
\multiput(1168.00,847.59)(0.762,0.482){9}{\rule{0.700pt}{0.116pt}}
\multiput(1168.00,846.17)(7.547,6.000){2}{\rule{0.350pt}{0.400pt}}
\multiput(1177.00,853.60)(0.627,0.468){5}{\rule{0.600pt}{0.113pt}}
\multiput(1177.00,852.17)(3.755,4.000){2}{\rule{0.300pt}{0.400pt}}
\multiput(1182.00,857.60)(0.774,0.468){5}{\rule{0.700pt}{0.113pt}}
\multiput(1182.00,856.17)(4.547,4.000){2}{\rule{0.350pt}{0.400pt}}
\multiput(1188.00,861.59)(0.933,0.477){7}{\rule{0.820pt}{0.115pt}}
\multiput(1188.00,860.17)(7.298,5.000){2}{\rule{0.410pt}{0.400pt}}
\multiput(1197.00,866.61)(0.462,0.447){3}{\rule{0.500pt}{0.108pt}}
\multiput(1197.00,865.17)(1.962,3.000){2}{\rule{0.250pt}{0.400pt}}
\multiput(1200.00,869.59)(0.798,0.485){11}{\rule{0.729pt}{0.117pt}}
\multiput(1200.00,868.17)(9.488,7.000){2}{\rule{0.364pt}{0.400pt}}
\put(1094.0,800.0){\usebox{\plotpoint}}
\multiput(1212.00,876.59)(0.692,0.488){13}{\rule{0.650pt}{0.117pt}}
\multiput(1212.00,875.17)(9.651,8.000){2}{\rule{0.325pt}{0.400pt}}
\put(1223,883.67){\rule{0.482pt}{0.400pt}}
\multiput(1223.00,883.17)(1.000,1.000){2}{\rule{0.241pt}{0.400pt}}
\multiput(1225.00,885.59)(0.626,0.488){13}{\rule{0.600pt}{0.117pt}}
\multiput(1225.00,884.17)(8.755,8.000){2}{\rule{0.300pt}{0.400pt}}
\put(1235,893.17){\rule{0.700pt}{0.400pt}}
\multiput(1235.00,892.17)(1.547,2.000){2}{\rule{0.350pt}{0.400pt}}
\multiput(1238.00,895.59)(0.762,0.482){9}{\rule{0.700pt}{0.116pt}}
\multiput(1238.00,894.17)(7.547,6.000){2}{\rule{0.350pt}{0.400pt}}
\multiput(1247.00,901.61)(0.909,0.447){3}{\rule{0.767pt}{0.108pt}}
\multiput(1247.00,900.17)(3.409,3.000){2}{\rule{0.383pt}{0.400pt}}
\multiput(1252.00,904.59)(0.599,0.477){7}{\rule{0.580pt}{0.115pt}}
\multiput(1252.00,903.17)(4.796,5.000){2}{\rule{0.290pt}{0.400pt}}
\multiput(1258.00,909.59)(0.710,0.477){7}{\rule{0.660pt}{0.115pt}}
\multiput(1258.00,908.17)(5.630,5.000){2}{\rule{0.330pt}{0.400pt}}
\multiput(1265.00,914.60)(0.627,0.468){5}{\rule{0.600pt}{0.113pt}}
\multiput(1265.00,913.17)(3.755,4.000){2}{\rule{0.300pt}{0.400pt}}
\multiput(1270.00,918.59)(0.762,0.482){9}{\rule{0.700pt}{0.116pt}}
\multiput(1270.00,917.17)(7.547,6.000){2}{\rule{0.350pt}{0.400pt}}
\put(1279,924.17){\rule{0.700pt}{0.400pt}}
\multiput(1279.00,923.17)(1.547,2.000){2}{\rule{0.350pt}{0.400pt}}
\multiput(1282.00,926.59)(0.645,0.485){11}{\rule{0.614pt}{0.117pt}}
\multiput(1282.00,925.17)(7.725,7.000){2}{\rule{0.307pt}{0.400pt}}
\put(1291,933.17){\rule{0.482pt}{0.400pt}}
\multiput(1291.00,932.17)(1.000,2.000){2}{\rule{0.241pt}{0.400pt}}
\multiput(1293.00,935.59)(0.692,0.488){13}{\rule{0.650pt}{0.117pt}}
\multiput(1293.00,934.17)(9.651,8.000){2}{\rule{0.325pt}{0.400pt}}
\put(1304,942.67){\rule{0.241pt}{0.400pt}}
\multiput(1304.00,942.17)(0.500,1.000){2}{\rule{0.120pt}{0.400pt}}
\multiput(1305.00,944.59)(0.560,0.488){13}{\rule{0.550pt}{0.117pt}}
\multiput(1305.00,943.17)(7.858,8.000){2}{\rule{0.275pt}{0.400pt}}
\multiput(1309.85,952.60)(-1.212,0.468){5}{\rule{1.000pt}{0.113pt}}
\multiput(1311.92,951.17)(-6.924,4.000){2}{\rule{0.500pt}{0.400pt}}
\put(1211.0,876.0){\usebox{\plotpoint}}
\put(393,258){\usebox{\plotpoint}}
\put(393,258){\circle*{24}}
\put(780,-400){\makebox(0,0){{\Large \bf Fig. 4}}}
\end{picture}\\
\end{document}